\documentclass[iop]{emulateapj}

\usepackage{amsmath}

\shorttitle{Double nuclei in NGC 908 and NGC 1187}
\shortauthors{Menezes \& Steiner}

\begin{document}

\title{Double nuclei in NGC 908 and NGC 1187}

\author{R. B. Menezes$^{1,2,3}$ and J. E. Steiner$^{1}$}

\affil{$^{1}$Instituto de Astronomia Geof\'isica e Ci\^encias Atmosf\'ericas, Universidade de S\~ao Paulo, Rua do Mat\~ao 1226, Cidade Universit\'aria, S\~ao Paulo, SP CEP 05508-090, Brazil; \\
$^{2}$Instituto de Pesquisa e Desenvolvimento, Universidade do Vale do Para\'iba, Av. Shishima Hifumi, 2911, S\~ao Jos\'e dos Campos, SP CEP 12244-000, Brasil; \\
$^{3}$Centro de Ci\^encias Naturais e Humanas, Universidade Federal do ABC, Brazil}

\email{roberto.menezes@ufabc.edu.br}

\begin{abstract}

We analyze optical data cubes of the nuclear regions of two late-type galaxies, NGC 908 and NGC 1187, obtained with the Integral Field Unit of the Gemini Multi-Object Spectrograph. Both data cubes show stellar structures consistent with double nuclei. The morphology of the line-emitting areas in the central region of NGC 1187 is also of a double nucleus, while the spatial morphology of the line-emitting areas in the data cube of NGC 908 is consistent with a circumnuclear asymmetric ring. The emission-line ratios of the nuclear spectra (and, actually, along the entire field of view) of both galaxies are characteristic of H II regions. In particular, based on its emission-line properties, the circumnuclear ring in NGC 908 can be identified as a star-forming ring. The observed spatial morphology of the stellar emission and also the differences in the properties of the stellar populations detected in the stellar nuclei of NGC 908 suggest that the most likely scenario to explain the double stellar nucleus in this object involves a minor merger, probably with a high mass ratio. On the other hand, considering the similar properties of the stellar populations in the stellar nuclei of NGC 1187, together with the stellar and gas kinematic properties, we conclude that the most likely scenario to explain the double stellar and gas nucleus in this galaxy involves the stellar and gas kinematics, in the form of a circular nuclear disk subject to perturbations.

\end{abstract}


\section{Introduction}

The nuclear region of galaxies is one of the most important structural components of these objects, as it can provide relevant information about their formation and evolution. Certain galaxies show unusual morphologies in their central area. A double nucleus is one of these peculiar morphologies that deserve special attention.

There are two well-established scenarios capable of explaining the presence of a double nucleus in a galaxy. The first one involves the occurrence of a merger. According to the hierarchical model of formation of galaxies, more massive galaxies are formed by many merger episodes involving less massive galaxies. Considering this scenario, it is actually expected to find galaxies in the merging phase, when their outer regions are already mixed, but their individual nuclei have still not coalesced. Since most galaxies appear to host a supermassive black hole (SMBH; Ferrarese \& Ford 2005), these noncoalesced nuclei may be associated with two SMBHs before their coalescence. Galaxy mergers bring gas to the inner parts of the merger remnants (e.g. Springel et al. 2005; Hopkins \& Hernquist 2009). This can lead to accretion onto the noncoalesced SMBHs, which may be detected as active galactic nuclei (AGNs). Indeed, hundreds of AGN pairs with projected separations higher than 10 kpc have already been detected (e.g. Hennawi et al. 2006; Myers et al. 2008; Green et al. 2010). On the other hand, only a few dual AGNs, with kiloparsec-scale separations, and even fewer AGN pairs with separations of hundreds or tens of parsecs have been observed (e.g. Rodriguez et al. 2006; Comerford et al. 2011; Fabbiano et al. 2011; Fu et al. 2011; Koss et al. 2011).

Double nuclei resulting from mergers have also been observed in objects with low or even without nuclear activity. \citet{nes15}, for example, analyzed Sloan Digital Sky Survey (SDSS) images and optical spectra of the elliptical galaxy MCG-01-12-005 (at the center of the X-ray cluster EXO 0422-086) and detected a double nucleus, without significant emission lines, with a projected separation of $3\arcsec$ ($\sim 2.4$ kpc). Using \textit{Hubble Space Telescope} (\textit{HST}) images and also data obtained with the Spectrograph for Integral Field Observations in the Near Infrared (SINFONI), at the \textit{Very Large Telescope} (VLT), \citet{maz16} detected a double nucleus at the center of the elliptical galaxy NGC 5419. The authors observed a low-luminosity AGN (LLAGN) at the brightest nucleus (coincident with the photometric center), but there is no clear evidence of AGN emission from the second nucleus, at a projected distance of $0\arcsec\!\!.25$ ($\sim 70$ pc). \citet{tha00} analyzed near-infrared spectra of NGC 5236 (M83), obtained with the VLT ISAAC spectrograph. The nuclear region of this object shows strong star formation and, based on the stellar velocity distribution, the authors detected a double nucleus, with a projected separation of $2\arcsec\!\!.7$ ($\sim 48$ pc) and one of the components coinciding with the photometric center. One interesting point to be mentioned is that, although there is not much information in the literature about this topic, mergers of two galaxies without a central SMBH would also result in a probable double stellar nucleus. In this case, the two nuclei in the merger remnant would represent the nuclear stellar clusters of the initial merging galaxies. Therefore, at least some nonactive double nuclei in galaxies may be explained by this model.  

The second scenario to explain the double nucleus in a galaxy assumes the existence of an eccentric stellar disk in the central region. M31 (the Andromeda galaxy) is certainly the most famous example of an object whose nuclear region is consistent with this model. \citet{lau93}, using \textit{HST} images, showed that this galaxy has a double nucleus, separated by a projected distance of $0\arcsec\!\!.49$ ($\sim 1.8$ pc). \citet{tre95} proposed that a model of a rotating eccentric stellar disk reproduces the main properties of the photometry and also of the stellar kinematics in the nuclear region of M31. According to this model, the faintest of the two stellar nuclei corresponds to the position of the SMBH, while the brightest nucleus is coincident with the apoapsis of the eccentric stellar disk. The higher brightness of the second nucleus can be explained by the fact that the stars tend to accumulate at the apoapsis, due to their lower velocities there. Later studies performed several refinements to this eccentric disk model \citep{pei03,sal04}. Using \textit{HST} images, \citet{lau96} detected a double nucleus at the center of NGC 4486B and suggested that such morphology could also be explained by the eccentric disk model of \citet{tre95}.

\begin{figure*}
\epsscale{1.0}
\plotone{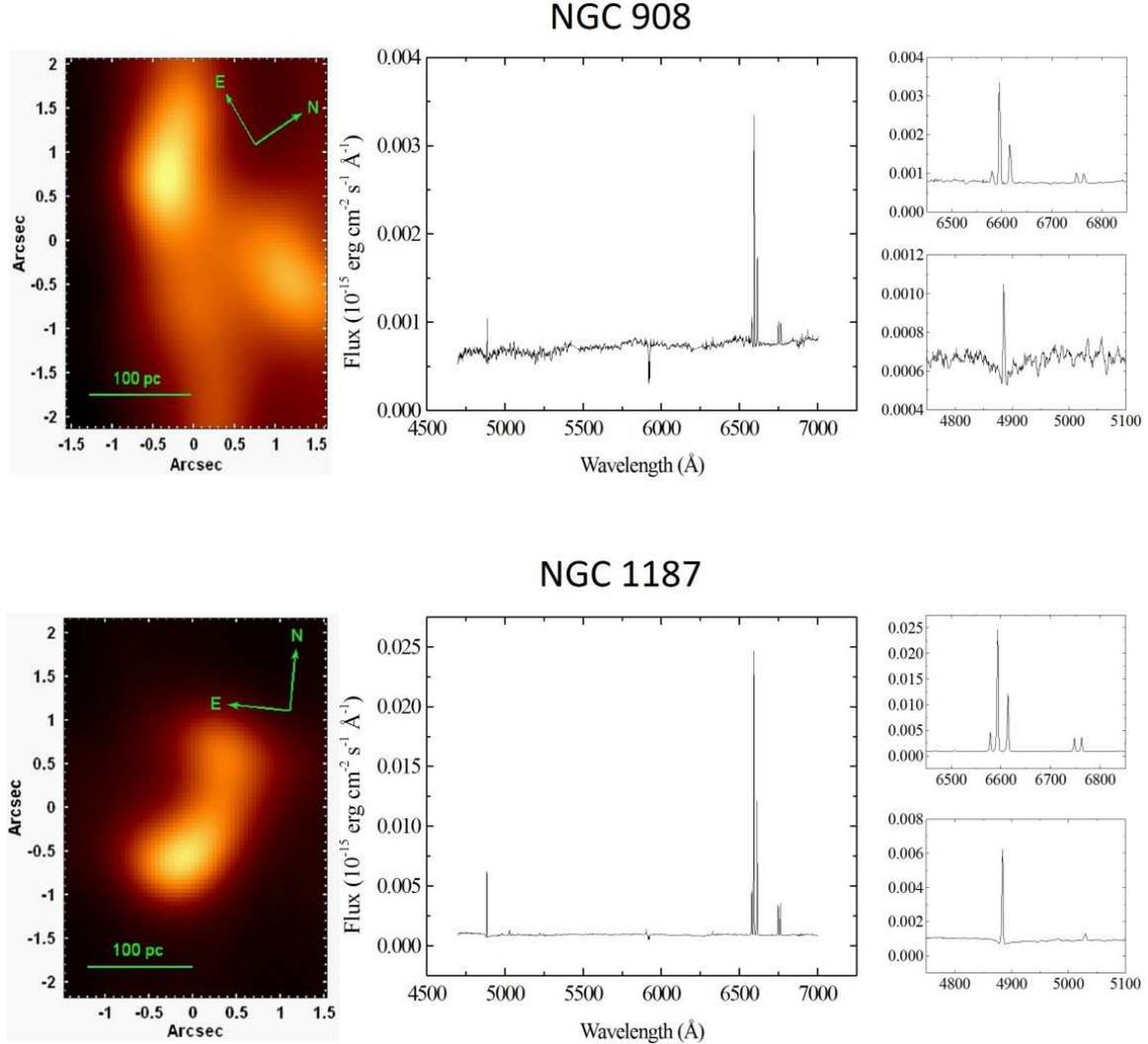}
\caption{Images of the collapsed treated data cubes of NGC 908 and NGC 1187, together with the corresponding average spectra. Magnifications of the blue and red regions of the average spectra are shown at right.\label{fig1}}
\end{figure*}

Therefore, one can say that the detection and analysis of double nuclei in galaxies can reveal much about the occurrence of mergers in these objects, which is certainly significant for studies of the hierarchical model of formation of galaxies, and also about the stellar kinematics around the central SMBHs, which is relevant for the determination of the SMBH masses. 

We are conducting the \textit{Deep IFS View of Nuclei of Galaxies} ($DIVING^{3D}$) survey of bright galaxies in the southern hemisphere. The purpose of this project is to observe, using optical 3D spectroscopy, the nuclear region of all galaxies in the southern hemisphere brighter than $B = 12.0$, which represents a total of 170 objects. Considering only the subsample of galaxies brighter than $B = 11.2$ (the ``$mini-DIVING^{3D}$'' sample, with a total of 57 objects), we found that NGC 613, NGC 908, and NGC 1187 show nuclear morphologies consistent with double nuclei. We intend to analyze the complex nuclear region of NGC 613 in a later work (da Silva et al. 2019 in preparation); in this paper we present the analysis of the data cubes of the nuclear regions of NGC 908 and NGC 1187, which are classified as SA(s)c and SB(r)c and are located at distances of 15.8 Mpc and 16.5 Mpc (NASA Extragalactic Database - NED), respectively. There is not much information in the literature about these objects, especially about their nuclear regions. 

In this work, we analyze the emission-line spectra, the stellar populations, and also the gas and stellar kinematics of the double nuclei in the data cubes of NGC 908 and NGC 1187. The observations, the data reduction, and the data treatment are described in Section 2. In Section 3, we analyze the emission-line properties of the data cubes. In Section 4, we present the analysis of the stellar populations. The stellar and gas kinematics are described in Section 5. Finally, we discuss our results and present our conclusions in Sections 6 and 7, respectively.

\section{Observations, Reduction and data treatment}

The observations of NGC 908 and NGC 1187 were taken on 2014 December 1 and 2014 December 23, respectively, using the Integral Field Unit (IFU) of the Gemini Multi-object Spectrograph (GMOS), at the Gemini-South telescope. Three 930 s exposures were taken for each object, with a $0\arcsec\!\!.2$ step spatial dithering. The GMOS/IFU was used in the one-slit mode. This configuration provides a science field of view (FOV), sampled by 500 fibers, with $3\arcsec\!\!.5 \times 5\arcsec\!\!.0$ and a sky FOV (at $1\arcmin$ from the science FOV), sampled by 250 fibers, with $5\arcsec\!\!.0 \times 1\arcsec\!\!.75$. We used the R831+G5322 grating, which resulted in a spectral coverage of 4675 - 7000 \AA~and in a spectral resolution of $R \sim 4300$. The seeing at the two nights of observation, estimated from the acquisition images (at 6300 \AA), was $\sim 0\arcsec\!\!.8$. For the flux calibration, the standard star LTT 2415 was observed on 2014 November 28.

The following baseline calibration images were obtained: bias, GCAL flat, twilight flat, and CuAr lamp. We reduced the data in IRAF environment, using the Gemini package. First, the data were trimmed and bias subtracted, and a cosmic-ray removal was applied, using the L. A. Cosmic routine \citep{van01}. After that, the spectra were extracted and a correction for gain variations along the spectral axis was performed, with response curves obtained from the GCAL flat images. A correction for gain variations (and for illumination patterns) along the FOV was also applied, with a response map provided by the twilight flat images. The spectra were then wavelength-calibrated, corrected for telluric absorptions, and flux-calibrated (taking into account the atmospheric extinction). Finally, the data cubes were constructed, with spatial pixels (spaxels) of $0\arcsec\!\!.05$.

The reduced data cubes of each object were corrected for the differential atmospheric refraction. This is a very important procedure, as this effect generates a displacement of the spatial structures along the spectral axis of the data cube, compromising any analysis to be performed. The three resulting data cubes of each galaxy were combined, in the form of a median, in order to correct for possible cosmic rays or even bad pixels that were not completely removed during the data reduction. After that, we applied a Butterworth spatial filtering \citep{gon02} to each image of the combined data cubes, in order to remove high spatial frequency structures. 

The GMOS/IFU data cubes usually show an ``instrumental fingerprint'' that appears as large vertical stripes across the images and that also has a characteristic low-frequency spectral signature. We do not know the origin of this instrumental fingerprint; however, our previous experiences showed that the Principal Component Analysis (PCA) Tomography technique (developed by our research group - Steiner et al. 2009) can efficiently isolate and remove this effect from the data cubes. Therefore, we used this procedure to perform the instrumental fingerprint removal in the data cubes of both galaxies analyzed in this work. Finally, we applied a Richardson-Lucy deconvolution \citep{ric72,luc74} to all the images of the data cubes, in order to improve the spatial resolution of the observations. This technique requires estimates of the point-spread functions (PSFs) of the data. Based on the data cube of the standard star LTT 2415, we verified that the FWHM of the PSF varies with the wavelength according to the following equation:  
\\
\begin{equation}
\resizebox{.8\hsize}{!}{$FWHM(\lambda) = FWHM_{ref}*(\frac{\lambda}{\lambda_{ref}})^{\alpha}$},
\end{equation}
\\
where $\alpha = -0.29$, $\lambda$ is the wavelength, FWHM($\lambda$) is the FWHM at $\lambda$, and FWHM$_{ref}$ is the FWHM at $\lambda_{ref}$. We also assumed that the PSF can be reproduced by a Moffat function, as determined from our extensive previous experiences. Since the data cubes of NGC 908 and NGC 1187 do not show point-like sources that could be used to estimate FWHM$_{ref}$ at a given $\lambda_{ref}$, we assumed that  FWHM$_{ref}$ is equal to the seeing value obtained from the acquisition images ($0\arcsec\!\!.8$), at $\lambda_{ref} = 6300$ \AA. We could not find higher-resolution images of these galaxies to perform a comparison to estimate the values of the final PSFs of the data cubes. However, considering the improvements usually provided by the Richardson-Lucy deconvolution, we conclude that the value of the FWHM of the PSF, at 6300 \AA, corresponding to the treated data cubes of NGC 908 and NGC 1187, is probably $\sim 0\arcsec\!\!.55$. More details about our data treatment methodology can be found in \citet{men14,men15, men18}. 

\begin{figure*}
\epsscale{0.85}
\plotone{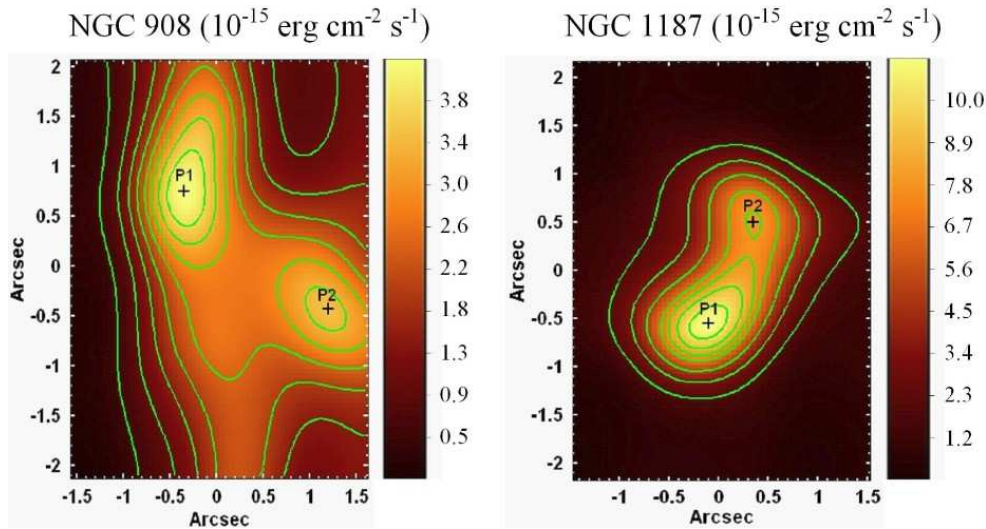}
\caption{Integrated flux images of the stellar data cubes of NGC 908 and NGC 1187, obtained with the synthetic stellar spectra provided by the spectral synthesis. The isocontours are shown in green and the positions of the stellar nuclei in each image are marked with crosses. For the analysis, the circumnuclear region is defined as the rest of the FOV, without taking into account the regions corresponding to the stellar nuclei.\label{fig2}}
\end{figure*}

Figure~\ref{fig1} shows images of the collapsed treated data cubes of NGC 908 and NGC 1187, together with the corresponding average spectra. The origin of the (x,y) coordinate system in all images in this paper was taken as the center of the FOV. The elongated structure seen in the image of the data cube of NGC 1187 is consistent with a double nucleus. On the other hand, the two nuclei are much more evident in the data cube of NGC 908. Both average spectra show weak absorption lines and prominent narrow emission lines, typical of H II regions.

\section{Analysis of the Emission-line spectrum}

The interstellar extinction in the data cubes of NGC 908 and NGC 1187, due to the Milky Way, was corrected using the extinction law of \citet{car89} and the following $A_V$ values (also from NED): $A_V = 0.069$ mag for NGC 908 and $A_V = 0.059$ mag for NGC 1187. We passed the data cubes to the rest frame, using the following redshift values, obtained from NED: $z = 0.005033$ for NGC 908 and $z = 0.004637$ for NGC 1187. We also resampled the spectra with $\Delta \lambda = 1$ \AA. In order to perform a reliable starlight subtraction, to analyze in detail the emission-line spectra, we applied a spectral synthesis to each spectrum of the data cubes, using the \textsc{STARLIGHT} software \citep{cid05}, which uses a combination of template spectra from a base to fit the stellar spectrum of a given object. One of the outputs obtained with this procedure is a synthetic stellar spectrum, corresponding to the fit of the observed spectrum. Since we applied the spectral synthesis to all the spectra of the data cubes, we were able to subtract the synthetic stellar spectra provided by the spectral synthesis from the observed ones, which resulted in data cubes containing only emission lines. The base of stellar population spectra used in this work is based on the Medium-resolution Isaac Newton Telescope Library of Empirical Spectra (MILES; S\'anchez-Bl\'azquez et al. 2006), with a spectral resolution of FWHM = $2.3$ \AA (similar to our spectral resolution of FWHM = $1.3$ \AA). This base has stellar population spectra with ages between $10^6$ yrs and $1.3 \times 10^{10}$ yrs and metallicities between $Z = 0.0001$ and $0.05$ ($Z = 0.02$ being the solar metallicity).

\begin{figure*}
\epsscale{1.15}
\plotone{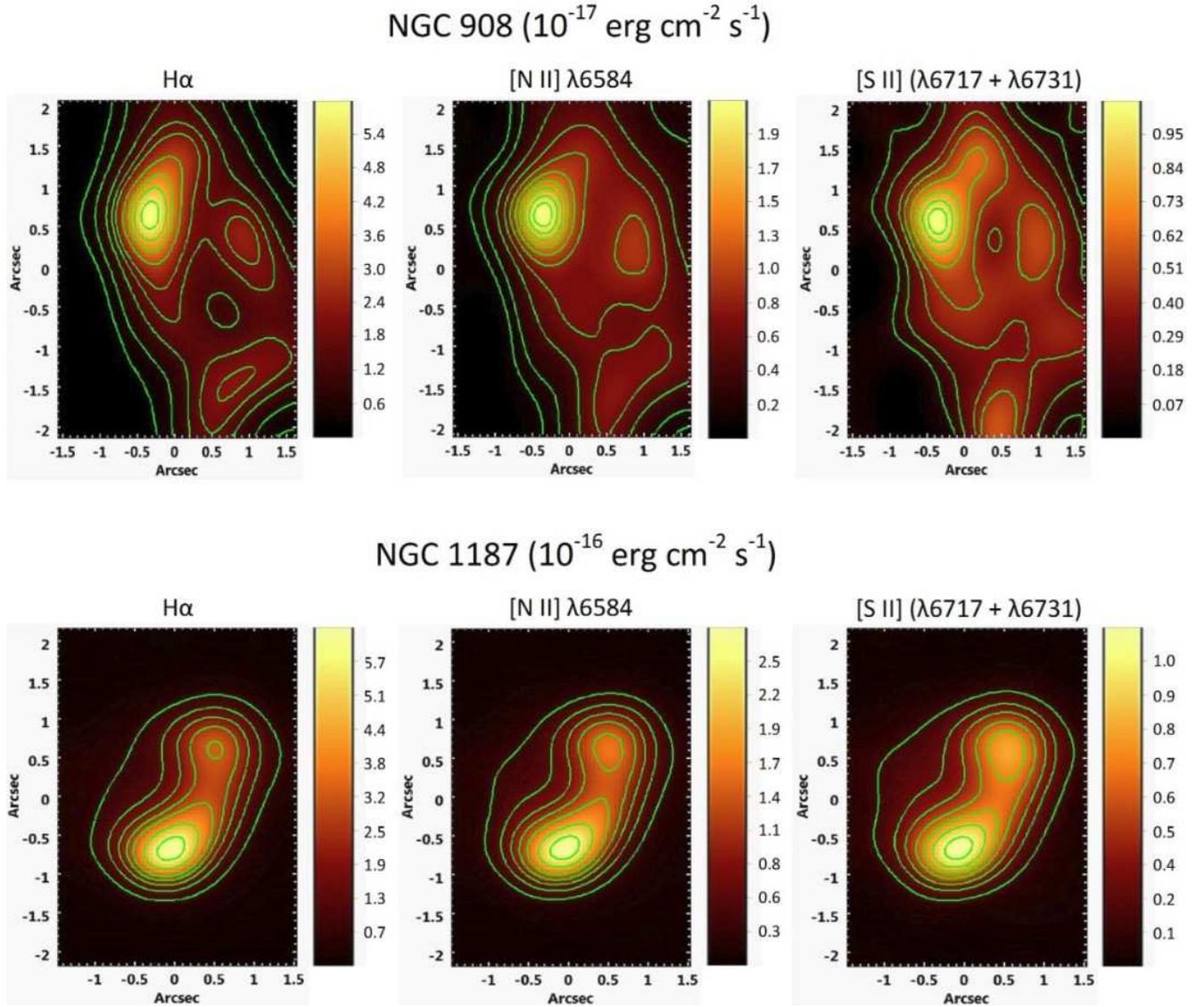}
\caption{Integrated flux images of the [N \textsc{II}] $\lambda6584$, [S \textsc{II}] $\lambda6717+\lambda6731$, and H$\alpha$ emission lines in the data cubes of NGC 908 and NGC 1187. The isocontours are shown in green.\label{fig3}}
\end{figure*}

We also constructed, for each galaxy, a stellar data cube with the synthetic spectra resulting from the spectral synthesis. Figure~\ref{fig2} shows integrated flux images of the stellar data cubes of NGC 908 and NGC 1187, revealing structures consistent with double stellar nuclei. The morphologies of these images are essentially the same as those of the images of the collapsed data cubes (Figure~\ref{fig1}). The integrated flux image of the stellar data cube of NGC 908 also shows elongated features, which are actually similar to tidal tails. The two stellar nuclei, in each data cube, were identified as P1 and P2, P1 being the brightest one. The position angle (PA) of the line joining P1 and P2 is $PA_{P1-P2}(NGC~908) = 111\degr \pm 2\degr$ and $PA_{P1-P2}(NGC~1187) = 163\degr \pm 3\degr$ in the data cubes of NGC 908 and NGC 1187, respectively. The projected distance between P1 and P2 in the data cubes of NGC 908 and NGC 1187 is $d_{P1-P2}(NGC~908) = 1\arcsec\!\!.94 \pm 0\arcsec\!\!.06$ and $d_{P1-P2}(NGC~1187) = 1\arcsec\!\!.14 \pm 0\arcsec\!\!.06$ , respectively.

Figure~\ref{fig3} shows integrated flux images, with isocontours, of the [N \textsc{II}] $\lambda6584$, [S \textsc{II}] $\lambda6717+\lambda6731$, and H$\alpha$ emission lines, obtained after the starlight subtraction. In the case of NGC 1187, all the images reveal a morphology compatible with a double nucleus, similar to that of the stellar component (Figure~\ref{fig2}). On the other hand, the emission-line images of the data cube of NGC 908 show a completely different structure, which can be identified as a circumnuclear asymmetric ring, with dimensions of $2\arcsec\!\!.6 \times 1\arcsec\!\!.3$ (200 pc $\times$ 100 pc, assuming a distance of 15.8 Mpc).

\begin{figure*}
\epsscale{0.88}
\plotone{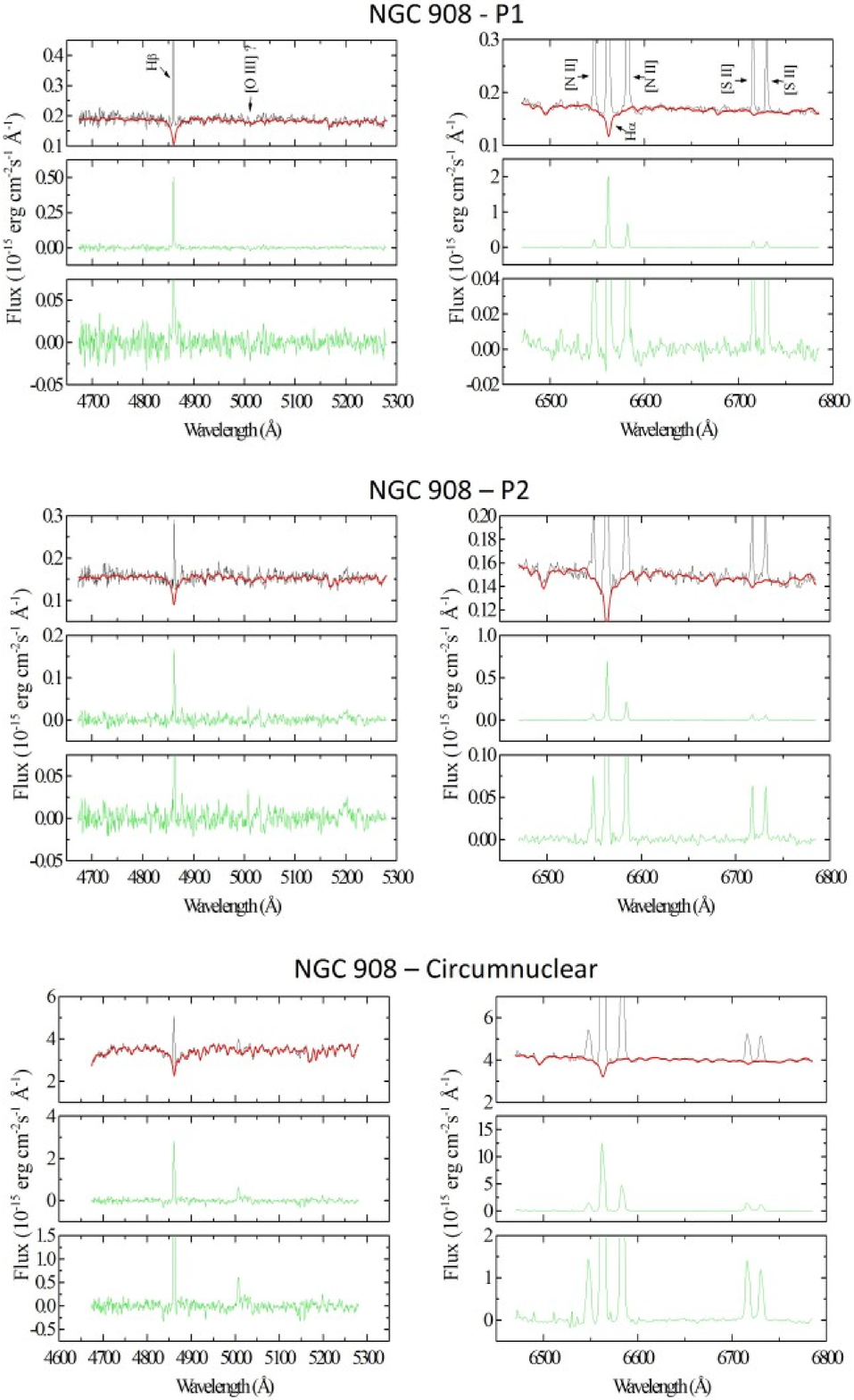}
\caption{Spectra extracted from circular regions, with a radius of $0\arcsec\!\!.28$, centered on P1 and P2 in the data cube of NGC 908. The fits provided by the spectral synthesis are shown in red and the fit residuals are shown in green.\label{fig4}}
\end{figure*}

\begin{figure*}
\epsscale{0.88}
\plotone{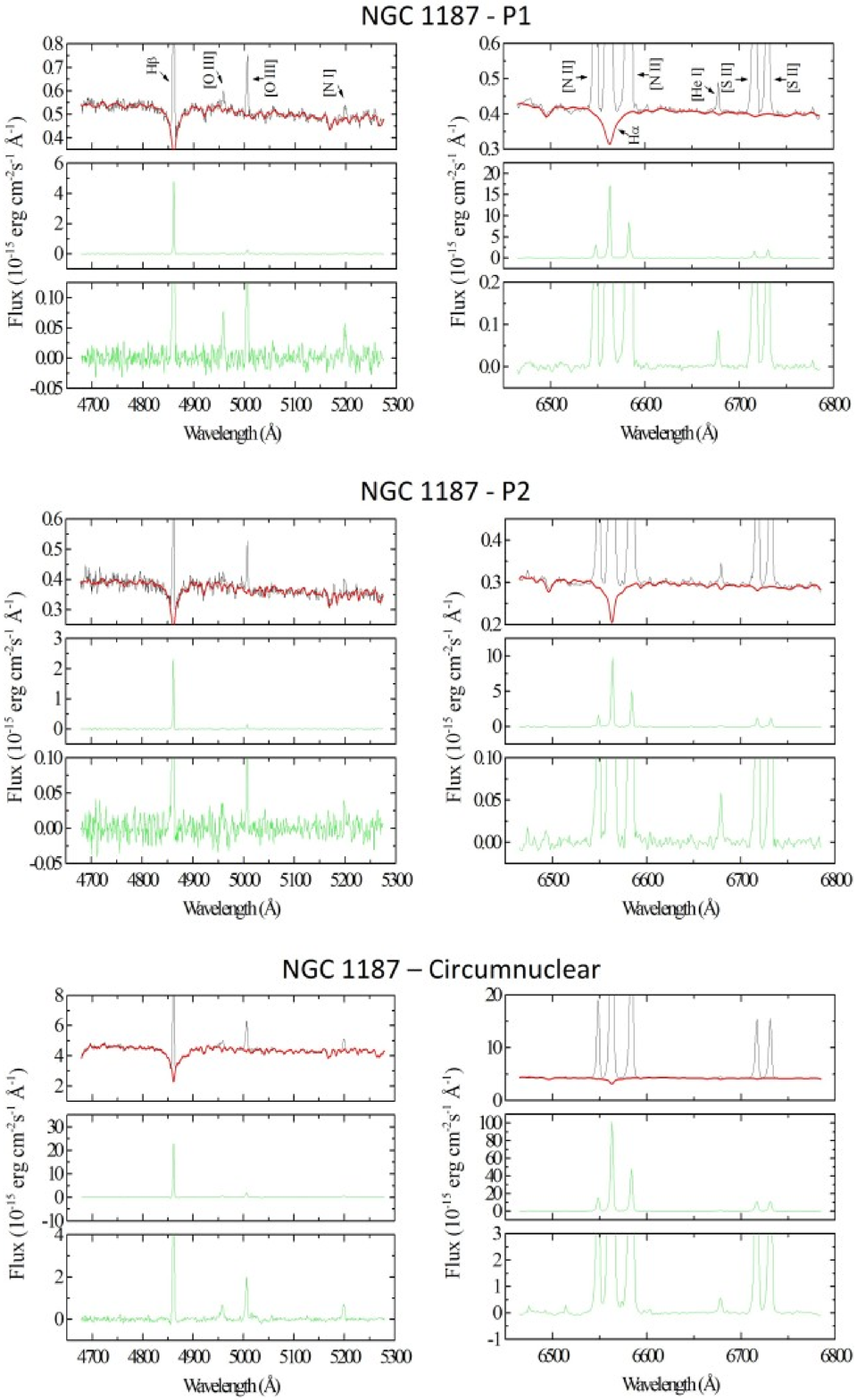}
\caption{Same as in Figure~\ref{fig4}, but for the data cube of NGC 1187\label{fig5}}
\end{figure*}

In order to analyze the properties of the two stellar nuclei in the data cubes of NGC 908 and NGC 1187, we extracted the spectra of circular regions, centered on each stellar nucleus of the data cubes before the starlight subtraction, with a radius of $0\arcsec\!\!.28$, corresponding to half of the FWHM of the PSF at the mean wavelength of the data cubes, which was estimated using equation 1. For each galaxy, we also subtracted the spectra of P1 and P2 from the total spectrum of the data cube, in order to obtain a spectrum representing the circumnuclear region. We applied the spectral synthesis to all the extracted spectra, which are shown in Figures~\ref{fig4} and~\ref{fig5}, together with the spectral synthesis fits and also the fit residuals. We can see that the absorption spectra were well reproduced by the fits. The emission-line spectra resulting from the starlight subtraction show strong and narrow H$\beta$, [N \textsc{II}] $\lambda\lambda 6584,6548$, H$\alpha$, and [S \textsc{II}] $\lambda\lambda 6717,6731$ lines. The [O \textsc{III}] $\lambda 5007$ emission line can be detected in the spectra of NGC 908 - P2, NGC 908 - Circumnuclear, and in all the spectra extracted from the data cube of NGC 1187, but only an upper limit for it can be estimated in the spectrum of NGC 908 - P1. The spectra of the data cube of NGC 1187 show also the [O \textsc{III}] $\lambda 4959$ and He \textsc{I} $\lambda 6678$ emission lines. 

\begin{table*}
\begin{center}
\caption{Balmer Decrement (before the Interstellar Extinction Correction), Emission-line Ratios, and Electron Densities Calculated for the Spectra Extracted from the Data Cubes of NGC 908 and NGC 1187.\label{tbl1}}
\begin{tabular}{ccccccc}
\hline
Nucleus & $\frac{H\alpha}{H\beta}$ & $\frac{[O \textsc{III}] \lambda 5007}{H\beta}$ & $\frac{[N \textsc{II}] \lambda 6584}{H\alpha}$ & $\frac{[S \textsc{II}] (\lambda 6716 + \lambda 6731)}{H\alpha}$ & $\frac{[S \textsc{II}] \lambda 6717}{[S \textsc{II}] \lambda 6731}$ & $n_e$ (cm$^{-3}$) \\
\hline
NGC 908 - P1 & 4.19 & $0.0254 \pm 0.0014$ & $0.320 \pm 0.013$ & $0.158 \pm 0.006$ & $1.02 \pm 0.09$ & $530 \pm 200$\\
NGC 908 - P2 & 3.88 & $0.11 \pm 0.07$ & $0.394 \pm 0.011$ & $0.219 \pm 0.014$ & $0.89 \pm 0.16$ & $880 \pm 620$ \\
NGC 908 - Circumnuclear & 5.62 & $0.27 \pm 0.07$ & $0.407 \pm 0.003$ & $0.232 \pm 0.004$ & $1.23 \pm 0.09$ & $90^{+130}_{-90}$ \\
NGC 1187 - P1 & 4.28 & $0.0761 \pm 0.0024$ & $0.450 \pm 0.009$ & $0.1919 \pm 0.0021$ & $0.863 \pm 0.017$ & $980 \pm 70$ \\
NGC 1187 - P2 & 5.25 & $0.069 \pm 0.004$ & $0.488 \pm 0.006$ & $0.222 \pm 0.003$ & $1.005 \pm 0.018$ & $560 \pm 40$ \\
NGC 1187 - Circumnuclear & 5.29 & $0.130 \pm 0.005$ & $0.464 \pm 0.008$ & $0.2266 \pm 0.0022$ & $1.010 \pm 0.011$ & $550 \pm 25$ \\
\hline
\end{tabular}
\end{center}
\end{table*}

\begin{table*}
\begin{center}
\caption{Median Stellar Ages and Metallicities, Extinction Values and $S/N$ Ratio Values in the Wavelength Range of 5910 - 6000\AA~Obtained from the Results Provided by the Spectral Synthesis Applied to the Spectra Extracted from the Data Cubes of NGC 908 and NGC 1187.\label{tbl2}}
\begin{tabular}{ccccc}
\hline
Nucleus & Median Age (yr) & Median Metallicity & $A_V$ & $S/N (cont)$ \\
\hline
NGC 908 - P1 & $(9.1 \pm 1.1) \times 10^6$ & $0.0216 \pm 0.0010$ & $1.49 \pm 0.06$ & 46 \\
NGC 908 - P2 & $(2.9 \pm 0.7) \times 10^8$ & $0.0240 \pm 0.0017$ & $1.41 \pm 0.04$ & 43 \\
NGC 908 - Circumnuclear & $(3.16 \pm 0.18) \times 10^8$ & $0.0293 \pm 0.0018$ & $1.573 \pm 0.020$ & 67 \\
NGC 1187 - P1 & $(9.1 \pm 1.8) \times 10^7$ & $0.011 \pm 0.005$ & $0.88 \pm 0.05$ & 67 \\
NGC 1187 - P2 & $(8.1 \pm 1.4) \times 10^7$ & $0.011 \pm 0.003$ & $0.96 \pm 0.07$ & 50 \\
NGC 1187 - Circumnuclear & $(2.04 \pm 0.07) \times 10^8$ & $0.0095 \pm 0.0014$ & $1.260 \pm 0.024$ & 69 \\
\hline
\end{tabular}
\end{center}
\end{table*}

\begin{figure*}
\epsscale{1.1}
\plotone{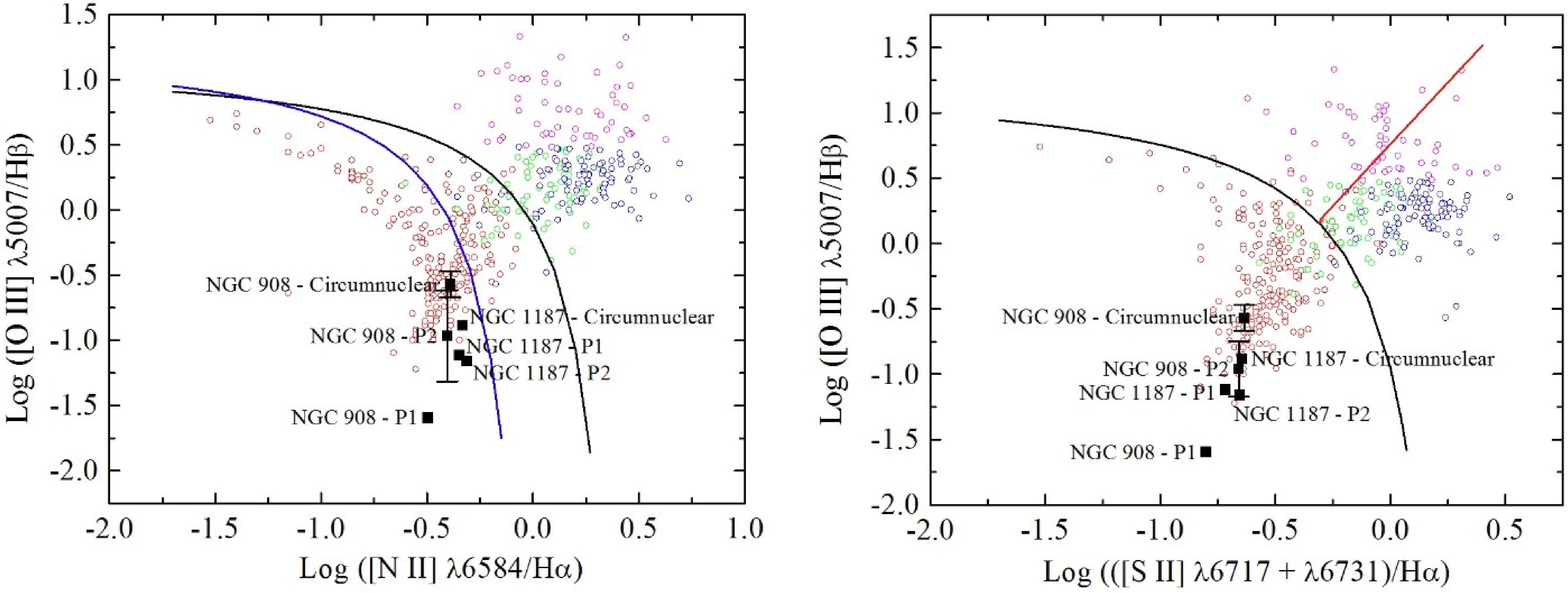}
\caption{Diagnostic diagrams with the points (filled black squares) corresponding to the nuclei and to the circumnuclear regions of NGC 908 and NGC 1187. The uncertainties are comparable to the sizes of the points without error bars. The other points represent the data analyzed by \citet{ho97}, with the red, green, blue, and magenta open circles corresponding to H II regions, transition objects, Low Ionization Nuclear Emission-Line Regions (LINERs), and Seyfert galaxies, respectively. The black curve represents the maximum limit for the ionization by a starburst, obtained by \citet{kew01}, the blue curve corresponds to the empirical division between H II regions and AGNs, obtained by \citet{kau03}, and the red curve represents the division between LINERs and Seyferts, determined by \citet{kew06}.\label{fig6}}
\end{figure*}

After the starlight subtraction, we calculated the Balmer decrement of all the extracted spectra. The results are shown in Table~\ref{tbl1}. Assuming that, in the absence of interstellar extinction, H$\alpha$/H$\beta$ = 2.86 (which is the value for the Case B of recombination, with a temperature of $10^4$ K and an electron density of $10^2$ cm$^{-3}$ - Osterbrock \& Ferland 2006) and also considering an extinction law of \citet{car89}, we used the obtained values of the Balmer decrement to correct the spectra for the interstellar extinction at the observed galaxies. We then calculated the integrated fluxes of H$\beta$, [N \textsc{II}] $\lambda 6548$, H$\alpha$, [S \textsc{II}] ($\lambda 6717 + \lambda 6731$), and [O \textsc{III}] $\lambda 5007$ and also the emission-line ratios [O \textsc{III}] $\lambda 5007$/H$\beta$, [N \textsc{II}] $\lambda 6584$/H$\alpha$, and [S \textsc{II}] ($\lambda 6716 + \lambda 6731$)/H$\alpha$ for all the extracted spectra. All the obtained values are shown in Table~\ref{tbl1}. Figure~\ref{fig6} shows the diagnostic diagrams of [N \textsc{II}] $\lambda 6584$/H$\alpha$ $\times$  [O \textsc{III}] $\lambda 5007$/H$\beta$ and [S \textsc{II}] ($\lambda 6716 + \lambda 6731$)/H$\alpha$ $\times$  [O \textsc{III}] $\lambda 5007$/H$\beta$, with the points representing NGC 908 - P1, NGC 908 - P2, NGC 908 - Circumnuclear, NGC 1187 - P1, NGC 1187 - P2, and NGC 1187 - Circumnuclear. Just to establish a comparison, we included, in the same graphs, the points corresponding to the objects analyzed by \citet{ho97}. The nuclei of NGC 908 and NGC 1187, as well as their circumnuclear regions, fall well in the branch of the H \textsc{II} regions, with a low ionization degree. Table~\ref{tbl1} also shows the values of the [S \textsc{II}] $\lambda 6717$/[S \textsc{II}] $\lambda 6731$ ratio and of the electronic density ($n_e$) obtained for all the extracted spectra. $n_e$ was calculated using the [S \textsc{II}] $\lambda 6717$/[S \textsc{II}] $\lambda 6731$ ratio and assuming a temperature of $10^4$ K \citep{ost06}. The $n_e$ values obtained for the spectra of P1 and P2 in the data cube of NGC 908 are compatible, at the 1$\sigma$ level. These densities are also compatible with the one obtained for the circumnuclear region of NGC 908, at the 2$\sigma$ level. However, considering the high uncertainties of such values, it is difficult to draw significant conclusions based on these results. The $n_e$ values calculated for the spectra of P1 and P2 in the data cube of NGC 1187 are considerably more precise and are not compatible, even at 3$\sigma$ level, indicating a real difference in the densities of these two regions. On the other hand, the densities obtained for P2 and for the circumnuclear region of NGC 1187 are compatible, at the 1$\sigma$ level. 

One important point to be discussed here is the fact that some of the derived $n_e$ values in Table~\ref{tbl1} are almost an order of magnitude larger than the $n_e$ assumed for the correction of the starlight-subtracted spectra for the interstellar extinction at the observed objects ($n_e = 10^2$ cm$^{-3}$). This could raise some questions about the reliability of the interstellar extinction correction applied. In order to analyze this point in further detail, we performed a simulation of a simple H II region with the \textsc{CLOUDY} software, last described by \citet{fer13}. In this simulation, we assumed a central OV5 star, with a luminosity of $3.98 \times 10^5$ L$_{\sun}$ and a temperature of 39000 K, and an initial $n_e$ value of $10^2$ cm$^{-3}$. We then repeated the simulation assuming all the $n_e$ values in Table~\ref{tbl1}. We observed a maximum variation of only $\sim$0.33\% from the original value of 2.86 for the H$\alpha$/H$\beta$ ratio, which is considerably lower than the uncertainties in this work. Therefore, we conclude that the interstellar extinction we applied to the starlight-subtracted spectra assuming H$\alpha$/H$\beta$ = 2.86 was adequate for our analyses.

\begin{figure*}
\epsscale{0.8}
\plotone{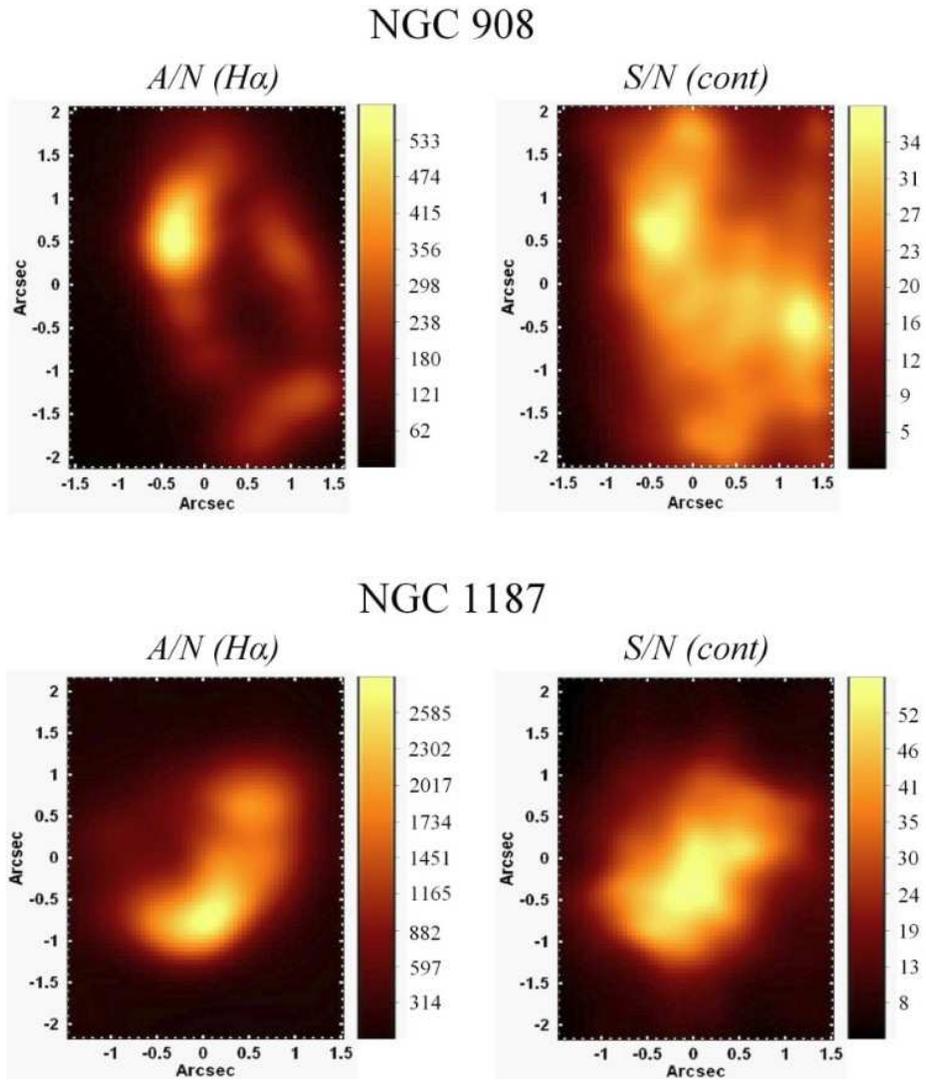}
\caption{Maps obtained for the $A/N$ ratio of the H$\alpha$ emission line and for the $S/N$ ratio, in the wavelength range 5910 - 6000 \AA, of the data cubes of NGC 908 and NGC 1187.\label{fig7}}
\end{figure*}

\section{Analysis of the stellar populations}

\begin{figure*}
\epsscale{1.1}
\plotone{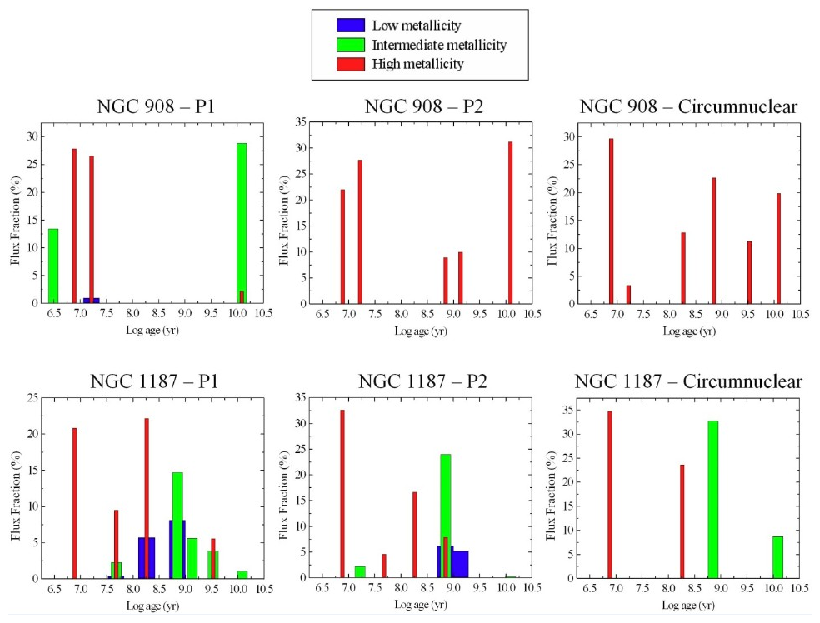}
\caption{Column graphs with the flux fractions corresponding to the stellar populations detected by the spectral synthesis of the spectra corresponding to the stellar nuclei and to the circumnuclear regions of NGC 908 and NGC1187. The blue, green, and red colors represent low ($Z = 0.0001$ and $0.0004$), intermediate ($Z = 0.004$ and $0.008$), and high ($Z = 0.02$ and $0.05$) metallicities, respectively. $Z_{\sun} = 0.02$ is the solar metallicity.\label{fig8}}
\end{figure*}

Besides a synthetic stellar spectrum, the spectral synthesis also provides, for each fitted spectrum, the values of the interstellar extinction ($A_V$) and of the flux fractions corresponding to all the stellar populations taken into account in the fit. Since the spectral synthesis was applied to the spectrum corresponding to each spaxel of the data cubes of NGC 908 and NGC 1187, one could construct $A_V$ maps and flux maps of the detected stellar populations, which would provide relevant information about the stellar content in the central regions of these galaxies. However, in order to perform such analysis, first of all, it is necessary to evaluate the signal-to-noise ($S/N$) ratio values of the stellar continuum along the FOV. We calculated the $S/N$ ratio, in the wavelength range of 5910 - 6000 \AA, of each spectrum of the data cubes of NGC 908 and NGC 1187. This wavelength range was chosen for the calculation because it is located nearly at the middle of our spectral coverage and also because it contains no significant spectral lines. The ``noise'' for this calculation was taken as the standard deviation of the values in this spectral region, after the starlight subtraction. The ``signal'', on the other hand, was taken as the average continuum flux in 5910 - 6000 \AA, before the starlight subtraction. The obtained $S/N$ ratio maps, shown in Figure~\ref{fig7}, show certain similarities to the images of the collapsed treated data cubes (Figure~\ref{fig1}). We verified that a minimum $S/N$ of 10 is necessary, in order to obtain, from the spectral synthesis, reliable information about the stellar populations. The $S/N$ ratio in the data cube of NGC 908 is lower than 10 in a nearly vertical region, east from the center of the FOV. At P1 and P2, the $S/N$ ratio reaches values of $\sim 37$. On the other hand, in the data cube of NGC 1187, the $S/N$ ratio values are lower than 10 in most of the regions located at projected distances from the center of the FOV greater than $1\arcsec\!\!.5$. The maximum values of $S/N$ are $\sim 57$, nearly at the center of the FOV. Therefore, the stellar population parameters provided by the spectral synthesis for large areas of the FOV are not reliable. We applied a spatial rebinning to the data cubes to obtain larger spaxels with higher $S/N$ values. However, even with spaxels with $0\arcsec\!\!.5$, large areas of the FOVs of the two analyzed data cubes remained with $S/N$ ratio values lower than 10. Considering that, we concluded that flux maps of the stellar populations detected by the spectral synthesis are not worthy of being included in this work. 

An alternative approach to analyze the stellar content in the central regions of NGC 908 and NGC 1187 is to use the spectra (analyzed in Section 3) of P1, P2, and of the circumnuclear region. The $S/N$ ratio values of such spectra (shown in Table~\ref{tbl2}) are higher than 10, and as a consequence, the associated stellar population parameters provided by the spectral synthesis can be considered reliable to be included in this work. Figure~\ref{fig8} shows column graphs with the flux fractions corresponding to the stellar populations detected by the spectral synthesis of the spectra of NGC 908 - P1, NGC 908 - P2, NGC 908 - Circumnuclear, NGC 1187 - P1, NGC 1187 - P2, and NGC 1187 - Circumnuclear. A detailed discussion of the results is given below. 

A simple visual inspection of the column graphs in Figure~\ref{fig8} may not reveal precisely how much the stellar populations in different spatial regions of the FOV of a data cube are consistent or not. Two parameters that can help to perform this kind of comparison are the median age ($T_{med}$) and the median metallicity ($Z_{med}$) of these graphs. In order to calculate such parameters, first of all, we determined, for each extracted spectrum, the cumulative flux fraction as a function of $T$ and $Z$ associated with each stellar population. Then, to obtain $T_{med}$, we calculated, using a cubic spline, the age of the stellar population with a cumulative flux corresponding to half of the total stellar flux of the spectrum. $Z_{med}$ was determined using an analogous procedure. These calculations were done in logarithmic scale. The uncertainties of these values were obtained using a Monte Carlo procedure. We constructed, for each spectrum, a histogram representing the spectral noise and fitted a Gaussian function to this histogram. After that, we obtained Gaussian distributions of random noise, with the same width of the Gaussian fitted to the histogram. These random distributions were added to the synthetic stellar spectrum corresponding to the fit of the original spectrum, provided by the spectral synthesis. We then applied the spectral synthesis to all the resulting spectra and calculated the corresponding $T_{med}$ and $Z_{med}$ values. The standard deviations of these values were taken as the uncertainties of $T_{med}$ and $Z_{med}$ of the original spectrum. The uncertainties of the $A_V$ values provided by the spectral synthesis were determined with this same Monte Carlo procedure. The values of $T_{med}$, $Z_{med}$, and $A_V$ obtained for each spectrum are given in Table~\ref{tbl2} and discussed below. It is worth emphasizing that $T_{med}$ and $Z_{med}$ are not used in this work as fundamental parameters describing the stellar populations, but merely as a mathematical criterion to determine how similar the stellar content from the extracted spectra is.

Our main purpose with the analysis presented in this section is to detect similarities and differences in the stellar content of different spatial regions in the data cubes of NGC 908 and NGC 1187. However, it is important to mention that the results provided by the spectral synthesis technique are usually affected by degeneracies. Therefore, such results must be handled with caution. One point in favor of the method is that, since the same technique was applied to all the extracted spectra, each of the obtained results is subject to the same kind of imprecisions. As a consequence, although the exact ages and metallicities of the detected stellar populations may not be totally reliable, the differences between them are likely to be. The same argument applies to the obtained $A_V$ values. Even so, the results provided by the spectral synthesis in this work should not be taken as conclusive and will only be used as possible evidences to evaluate the most likely scenarios to explain the morphologies of the double nuclei in NGC 908 and NGC 1187.

\subsection{NGC 908}

Figure~\ref{fig8} shows that most of the stellar flux ($\sim 68$\%) in NGC 908 - P1 is due to young stellar populations, with ages of $\sim 10^6$ and $\sim 10^7$ yrs and with intermediate ($Z = 0.004$ and $0.008$) and high ($Z = 0.02$ and $0.05$) metallicities. There is also an old stellar population, associated with $\sim 29$\% of the total stellar flux, with an age of $\sim 10^{10}$ yrs and with an intermediate metallicity. The graph of NGC 908 - P2 shows significant differences. A large flux fraction ($\sim 50$\%) is still due to young stellar populations, with ages of $\sim 10^7$ yrs. However, the very young ($\sim 10^6$ yrs) stellar population with intermediate metallicity observed in the spectrum of NGC 908 - P1 was not detected in the spectrum of NGC 908 - P2. This last spectrum also shows high-metallicity stellar populations, corresponding to a flux fraction of $\sim 19\%$, with ages of $\sim 10^9$ yrs, which do not appear in the spectrum of NGC 908 - P1. Similarly to what was observed for NGC 908 - P1, a large flux fraction ($\sim 31$\%) of the spectrum of NGC 908 - P2 is due to an old stellar population, with an age of $\sim 10^{10}$ yrs. However, such a stellar population has a high metallicity, unlike the one observed in NGC 908 - P1, which has an intermediate metallicity. The graph of the spectrum of NGC 908 - Circumnuclear shows many similarities to the one of the spectrum of NGC 908 - P2, such as the presence of young ($\sim 10^7$ yrs), intermediate-age ($\sim 10^9$ yrs), and old ($\sim 10^{10}$ yrs) stellar populations and the complete absence of low- and intermediate-metallicity populations. The main difference between the graphs of NGC 908 - P2 and NGC 908 - Circumnuclear is the fact that the flux due to intermediate-age stellar populations is more significant in the latter.

\begin{figure*}
\epsscale{1.15}
\plotone{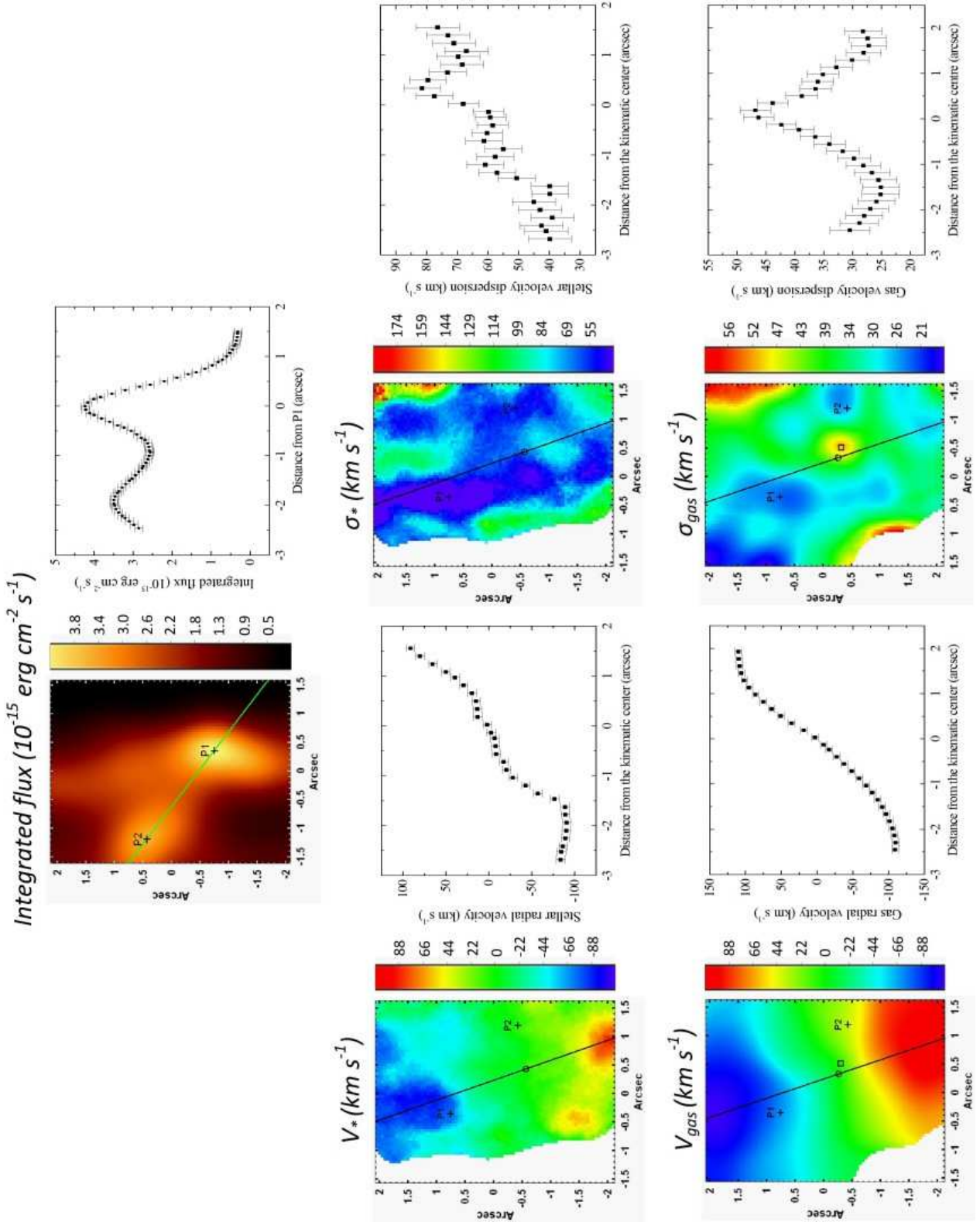}
\caption{Stellar and gas kinematic maps of the data cube of NGC 908. The black line in the stellar kinematic maps represents the line of nodes of the $V_*$ map. Similarly, the black line in the gas kinematic maps represents the line of nodes of the $V_{gas}$ map. The kinematic centers of the $V_*$ and $V_{gas}$ maps are marked with open circles and the position of the $\sigma_{gas}$ peak is marked with an open square. The positions of P1 and P2 are marked with crosses. The curves extracted, from all kinematic maps, along the lines of nodes of the $V_*$ and $V_{gas}$ maps are shown at the sides of the corresponding maps. The areas where the $A/N$ ratio of the H$\alpha$ line or the $S/N$ ratio, in the wavelength range 5910 - 6000 \AA, are lower than 10 were masked in the maps. The integrated flux images of the stellar data cubes of NGC 908 and NGC 1187 (Figure~\ref{fig2}), together with curves extracted along the lines connecting P1 and P2 in each data cube, are also shown.\label{fig9}}
\end{figure*}

\begin{figure*}
\epsscale{1.15}
\plotone{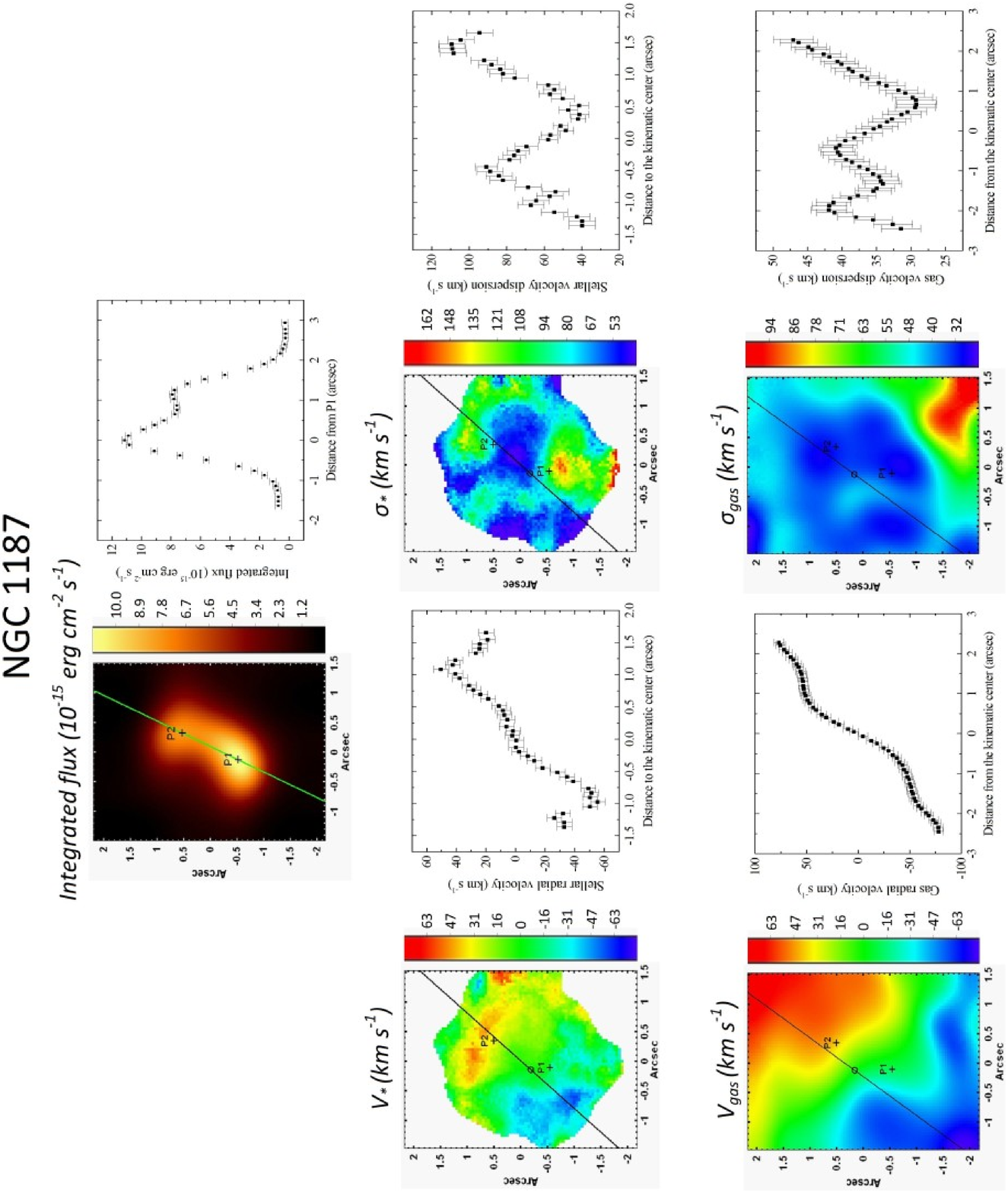}
\caption{Same as in Figure~\ref{fig9}, but for the data cube of NGC 1187.\label{fig10}}
\end{figure*}

The analysis involving the parameters $T_{med}$ and $Z_{med}$ reveals a significant difference between the stellar content in NGC 908 - P1 and NGC 908 - P2, as the $T_{med}$ values of these stellar nuclei are not compatible, even at the 3$\sigma$ level, although the $Z_{med}$ values are compatible, at the 1$\sigma$ level. The $T_{med}$ values of NGC 908 - P1 and NGC 908 - Circumnuclear are not compatible, even at the 3$\sigma$ level, and the corresponding $Z_{med}$ values are compatible, at the 3$\sigma$ level, which also indicates a considerable difference between the stellar content in these regions. On the other hand, the analysis shows a similarity between the stellar populations in NGC 908 - P2 and NGC 908 - Circumnuclear, as the $T_{med}$ and $Z_{med}$ values of these areas are compatible, at 1$\sigma$ and 2$\sigma$ levels, respectively. The interstellar extinction along the FOV of the data cube of NGC 908 does not change considerably, as the $A_V$ values obtained for NGC 908 - Circumnuclear and NGC 908 - P1 are compatible, at 1$\sigma$ level, and the ones obtained for NGC 908 - Circumnuclear and NGC 908 - P2 are also compatible, at the 3$\sigma$ level. Similarly, the $A_V$ values determined for NGC 908 - P1 and NGC 908 - P2 are compatible, at the 1$\sigma$ level.

\subsection{NGC 1187}

The column graph of NGC 1187 - P1 in Figure~\ref{fig8} shows that the observed stellar flux is due to stellar populations with a considerable variety of ages and metallicities. More than $90\%$ of the flux is due to young ($\sim10^7$ yrs) and intermediate-age ($\sim 10^8$ and $10^9$ yrs) stellar populations, with low ($Z = 0.0001$ and $0.0004$), intermediate, and high metallicities. The graph of NGC 1187 - P2 reveals essentially the same behavior, although with less variety of ages. The graph of NGC 1187 - Circumnuclear, on the other hand, shows certain differences, such as the absence of flux due to low-metallicity stellar populations and even less variety of ages.

Based on the values in Table~\ref{tbl2}, the stellar content in P1 and P2 appears to be very similar, as the values of $T_{med}$ and $Z_{med}$ in these regions are compatible, at the 1$\sigma$ level. The $Z_{med}$ value of NGC 1187 - Circumnuclear is compatible, at the 1$\sigma$ level, with the ones of NGC 1187 - P1 and NGC 1187 - P2. However, the $T_{med}$ value of NGC 1187 - Circumnuclear is not compatible with the ones of the two stellar nuclei, even at the 3$\sigma$ level, revealing a measurable difference between the stellar populations in these regions. The changes of the interstellar extinction along the FOV of the data cube of NGC 1187 are more significant than in the case of the data cube of NGC 908. The $A_V$ values obtained for NGC 1187 - P1 and NGC 1187 - P2 are compatible, at the 1$\sigma$ level; however, these two values are not compatible with the one determined for NGC 1187 - Circumnuclear, even at the 3$\sigma$ level.

\section{Analysis of the stellar and gas kinematics}

The spectral synthesis, described in Section 3, also provides the values of the stellar radial velocity ($V_*$) and of the stellar velocity dispersion ($\sigma_*$). Therefore, in order to analyze these parameters, we constructed, for each galaxy, maps of the values of $V_*$ and $\sigma_*$, using the results obtained with the spectral synthesis. The obtained maps, together with the curves extracted along the lines of nodes (corresponding to the lines connecting the points with maximum and minimum values of $V_*$) of the $V_*$ maps, are shown in Figures~\ref{fig9} and~\ref{fig10} and the results are discussed below. In these figures, we also show, for comparison, the integrated flux images of the stellar data cubes of NGC 908 and NGC 1187 (the same as in Figure~\ref{fig2}), together with curves extracted along the line joining P1 and P2. We can see that, for both galaxies, the integrated stellar flux values at P1 and P2 are not compatible, even at the 3$\sigma$ level, indicating that P1 is indeed the brightest of the two nuclei in the wavelength range of the data cubes. We assumed that the center of each $V_*$ map is given by the point, along the line of nodes, where the velocity is equal to the mean between the minimum and maximum velocities in the map. This mean velocity was subtracted from the corresponding $V_*$ map, so all the $V_*$ values in the maps in Figures~\ref{fig9} and~\ref{fig10} are given relative to the centers of the maps. The $\sigma_*$ values were corrected taking into account the instrumental spectral resolution and also the spectral resolution of the stellar population base used to perform the spectral synthesis. The uncertainties of $V_*$ and $\sigma_*$ were obtained with a Monte Carlo procedure (similar to the one used to estimate the uncertainties of $T_{med}$ and $Z_{med}$ - see Section 4), which included: constructing a histogram representing the spectral noise; fitting a Gaussian function to the histogram; creating Gaussian distributions of random noise (with the same width of the Gaussian fitted to the histogram); adding the random noise to the synthetic stellar spectrum provided by the spectral synthesis. Based on the $S/N$ ratio (Figure~\ref{fig7}) maps and also on the uncertainties estimated for $V_*$ and $\sigma_*$, we concluded that a minimum $S/N$ ratio of 10 is necessary, in order to obtain reliable values for these kinematic parameters. Therefore, the regions in the stellar kinematic maps of NGC 908 and NGC 1187 with $S/N$ ratio values lower than 10 were masked.

In order to analyze the gas kinematics in the data cubes of NGC 908 and NGC 1187, first of all, we evaluated the values of the amplitude-to-noise ($A/N$) ratio of the H$\alpha$ emission line (which is the most intense line in the data cubes), after the starlight subtraction, along the FOV. The ``noise'' for this calculation was taken as the standard deviation of the values in a spectral region adjacent to the H$\alpha$ emission line (but without other emission lines). The obtained $A/N$ ratio maps are shown in Figure~\ref{fig7}. The morphologies of the two maps are very similar to the ones of the integrated flux images of H$\alpha$ (Figure~\ref{fig3}), as expected. In the case of NGC 908, the $A/N$ ratio of H$\alpha$ is higher than 10 in most of the FOV, reaching values of $\sim 590$ close to P1. In the same map, there is a small area, southwest from P1, where the $A/N$ values are lower than 10, with minimum values of $\sim 4$. On the other hand, in the data cube of NGC 1187, the $A/N$ ratio of H$\alpha$ is higher than 10 along the entire FOV, with minimum values of $\sim 30$, in the most peripheral areas, and maximum values of $\sim 2870$, close to P1.

We then fitted a Gaussian function to the H$\alpha$ emission line of each spectrum of the data cubes, after the starlight subtraction. We verified that such fits provide reliable values for the radial velocity ($V_{gas}$) and for the velocity dispersion ($\sigma_{gas}$) of the gas, for $A/N$ ratio values higher than 10. Since this procedure was applied to each spectrum of the data cubes, we obtained maps of the values of $V_{gas}$ and $\sigma_{gas}$, which are shown in Figures~\ref{fig9} and~\ref{fig10}, together with the curves extracted along the lines of nodes of the $V_{gas}$ maps. The discussion of the results is given below. In the case of NGC 908, we masked the $V_{gas}$ and $\sigma_{gas}$ values in areas with $A/N$ ratio values of H$\alpha$ lower than 10. We assumed that the center of each $V_{gas}$ map corresponds to the point, along the line of nodes, with a velocity equal to the mean between the minimum and maximum velocities in the map. Such mean velocity was subtracted from the $V_{gas}$  map. The $\sigma_{gas}$ values were corrected for the instrumental resolution. The uncertainties of the values were estimated with the same Monte Carlo procedure used to obtain the uncertainties of $V_*$ and $\sigma_*$. The main difference here is that the Gaussian distributions of random noise used to apply this process (as described above) were sequentially added to the Gaussian fits of the H$\alpha$ emission line (which resulted in ``noisy simulated emission lines'').

\subsection{NGC 908}

Figure~\ref{fig9} shows that the $V_*$ map and the corresponding $V_*$ curve of NGC 908 are somewhat irregular but reveal a rotation pattern. Despite the observed irregularities, we can estimate an amplitude of $\sim 100$ km s$^{-1}$ for the $V_*$ curve. The PA of the line of nodes is $PA_* = 77\degr \pm 2\degr$, which is not compatible with the PA of the line joining P1 and P2 ($PA_{P1-P2}(NGC~908) = 111\degr \pm 2\degr$), even at 3$\sigma$ level. The $\sigma_*$ map and curve of this galaxy show that the values are lower in the area closer to the line of nodes of the $V_*$ map. The characteristics of the $V_*$ and $\sigma_*$ maps suggest the presence of a cold rotating stellar disk. It is worth emphasizing that the word ``cold'' here is used only to indicate that $\sigma_*$ is lower along the disk than in surrounding regions. The term is not related to specific values of $\sigma_*$. A comparison between the $\sigma_*$ map and the integrated flux images of the [N \textsc{II}] $\lambda6584$, [S \textsc{II}] $\lambda6717+\lambda6731$, and H$\alpha$ emission lines in the data cube of NGC 908 (Figure~\ref{fig3}) reveals a similarity in the morphologies of these images, with the $\sigma_*$ values lower along the extension of the asymmetric emitting circumnuclear ring described before (for more details, see Section 6). The $\sigma_*$ map does not show any significant $\sigma_*$ peak in the region along which the rotation pattern takes place.

The $V_{gas}$ map and the corresponding $V_{gas}$ curve, in Figure~\ref{fig9}, of NGC 908 reveal a pattern consistent with rotation, the $V_{gas}$ curve having an amplitude of $\sim 110$ km s$^{-1}$. The PA of the line of nodes ($PA_{gas} = 76\degr \pm 2\degr$) is compatible, at the 1$\sigma$ level, with $PA_*$. The centers of the $V_{gas}$ and $V_*$ maps are compatible, at the 2$\sigma$ level, assuming an uncertainty of $0\arcsec\!\!.1$ for each of these positions. The $\sigma_{gas}$ map of NGC 908 shares certain morphological similarities to the $\sigma_*$ map of this object, although the $\sigma$ values in these two maps are considerably different. As observed in the $\sigma_*$ map, $\sigma_{gas}$ is lower in the area closer to the line of nodes of the $V_{gas}$ map, which suggests the presence of a cold rotating gas disk, corotating with the probable stellar disk. A $\sigma_{gas}$ peak of $\sim 45$ km s$^{-1}$ can also be noted. Such a peak is not exactly coincident with the $V_{gas}$ map center, but these positions, with uncertainties of $0\arcsec\!\!.05$ and $0\arcsec\!\!.1$, respectively, are compatible, at the 1$\sigma$ level. There is a morphological similarity (also observed in the $\sigma_*$ map) between the $\sigma_{gas}$ map and the integrated flux images of the [N \textsc{II}] $\lambda6584$, [S \textsc{II}] $\lambda6717+\lambda6731$, and H$\alpha$ emission lines (Figure~\ref{fig3}), with the $\sigma_{gas}$ values lower along the extension of the observed asymmetric emitting circumnuclear ring.

\subsection{NGC 1187}

Although a significant part of the $V_*$ map of NGC 1187 in Figure~\ref{fig10} was masked, we can identify a rotation pattern in the inner region. The amplitude of the $V_*$ curve is $\sim 50$ km s$^{-1}$. The PA of the line of nodes is $PA_* = 147\degr \pm 3\degr$, which is compatible, at the 3$\sigma$ level, with the PA of the line joining P1 and P2 ($PA_{P1-P2}(NGC~1187) = 163\degr \pm 3\degr$). The $\sigma_*$ map is considerably irregular, and as a consequence, it is difficult to confirm whether the observed structures are real or a consequence of instabilities of the fitting procedure.

The $V_{gas}$ map of NGC 1187 in Figure~\ref{fig10} shows a slightly more complex morphology. There is certainly some indication of rotation, but a few disturbances, especially within a radius of $1\arcsec\!\!.5$ from the kinematic center, can also be seen. In the $V_{gas}$ curve, there is a typical rotation pattern within a radius of $\sim 1\arcsec\!\!.5$ from the kinematic center, with an amplitude of $\sim 50$ km s$^{-1}$. Such a pattern is consistent with what was observed in the $V_*$ map and curve of this galaxy. However, at larger radii, there is a change in such a pattern. This difference between the inner and outer kinematic behavior may indicate the presence of two distinct gas kinematic phenomena in the central region of this galaxy. The PA of the line of nodes obtained for the $V_{gas}$ map of NGC 1187 ($PA_{gas} = 153\degr \pm 2\degr$) is compatible, at the 2$\sigma$ level, with $PA_*$. Considering an uncertainty of $0\arcsec\!\!.1$ for the positions of the centers of the $V_{gas}$ and $V_*$ maps, we conclude that their positions are compatible, at the 1$\sigma$ level. The $\sigma_{gas}$ map of NGC 1187 shows certain similarities to that obtained for NGC 908, with lower values in the area closer to the line of nodes, suggesting the presence of a cold rotating gas disk. The main difference here is that there is no significant $\sigma_{gas}$ peak close to the kinematic center.

\section{Discussion}

\subsection{NGC 908}

The analyzed data cubes revealed very peculiar morphological features. It is clear from Figure~\ref{fig2} that both galaxies show a double stellar nucleus. In the case of NGC 908, we can even see elongated structures very similar to tidal tails (Knierman et al. 2012; Mullan et al. 2011; Struck \& Smith 2012; Knierman et al. 2013; Mulia et al. 2015; Wen \& Zheng 2016). One natural explanation, then, to the double stellar nucleus in NGC 908 is that it is the result of a merger. In that case, the two observed stellar concentrations are the noncoalesced stellar nuclei of the galaxies involved in the merger. The differences detected by the spectral synthesis in the stellar populations of P1 and P2 (Figure~\ref{fig8}), together with the fact that the values of $T_{med}$ of such stellar populations are not compatible, even at the 3$\sigma$ level, is consistent with the merger hypothesis, as there is no reason for the two merging galaxies to have stellar populations with similar ages. The results provided by the spectral synthesis are subject to degeneracies. As mentioned before, the fact that the same analysis technique was applied to all the extracted spectra suggests that the obtained results are all affected by the same kind of imprecisions, and as a consequence, the observed differences between the stellar population parameters may be reliable. Even so, we performed an additional test to evaluate that. Using the Starlight software, we fitted again the extracted spectrum of NGC 908 - P2, but taking into account only the stellar populations detected by the spectral synthesis applied to the spectrum of NGC 908 - P1. The $\chi^2$ value of this new fit ($\chi^2_{P2}(base~P1) = 2387$) was significantly larger than the $\chi^2$ obtained with the original fit ($\chi^2_{P2} = 2337$). We repeated the same procedure with the spectrum of NGC 908 - P1 and fitted it taking into account only the stellar populations detected by the spectral synthesis applied to the spectrum of NGC 908 - P2. Again, the $\chi^2$ value of the new fit ($\chi^2_{P1}(base~P2) = 1988$) was larger than the original one ($\chi^2_{P1} = 1936$). These results support the hypothesis that the stellar populations in NGC 908 - P1 and in NGC 908 - P2 are indeed considerably different. The discrepancy between $PA_*$ and $PA_{P1-P2}$ in the data cube of NGC 908 is evidence that the origin of the double stellar nucleus in this galaxy is not directly related to the possible nuclear stellar disk revealed by the $V_*$ map (see more details below). We cannot prove the existence of a rotating nuclear stellar disk in this object without a dynamical modeling. If the observed pattern is indeed due to a disk, its orientation is such that NGC 908 - P1 and NGC 908 - P2 cannot both lie in the disk plane.

\begin{figure*}
\epsscale{0.85}
\plotone{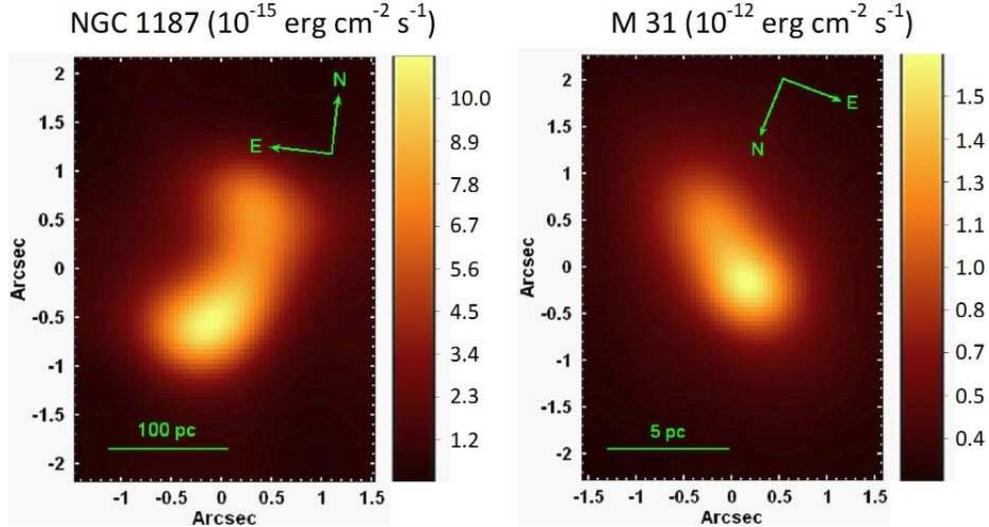}
\caption{Comparison of the integrated flux images obtained for the stellar data cubes of NGC 1187 and M31.\label{fig11}}
\end{figure*}

As mentioned before, NGC 908 is classified as SA(s)c, which is not an expected outcome of a major merger of galaxies. Therefore, if the merger hypothesis to explain the double stellar nucleus in this object is indeed correct, then we believe that the occurrence of a minor merger is the most likely scenario. A smaller (possibly satellite) galaxy being cannibalized by NGC 908 is a possibility. One interesting point to be discussed here is related to the star-forming activity in this galaxy. Although NGC 908 does not show strong star formation in its nuclear region, the emission-line spectrum along the entire extension of the circumnuclear asymmetric emitting ring observed in the H$\alpha$, [N \textsc{II}] $\lambda 6584$, and [S \textsc{II}] $\lambda 6717 + \lambda 6731$ integrated flux images is characteristic of H II regions. Therefore, this structure can be identified as a star-forming ring, probably composed of multiple H II regions. It is well known that galaxy mergers can lead to inflows of gas toward the center of the merger remnant and also to starbursts in that area (e.g. Barnes \& Hernquist 1991; Mihos \& Hernquist 1996). \citet{cox08} proposed that the merger-driven star formation depends on the merger mass ratio, being larger for major mergers (smaller mass ratios) and smaller for minor mergers (larger mass ratios). Similarly, \citet{tan96} proposed that the SMBH binary formed by a merger with a satellite galaxy results in mild star-forming activity, with several circumnuclear H II region clumps and possibly hot-spot nuclei, when the mass of the secondary SMBH (located at the satellite galaxy) is considerably lower ($\sim 10 \%$) than the mass of the primary SMBH. On the other hand, the nuclear star formation is much more significant when the secondary SMBH has a mass comparable to half of the primary SMBH mass. Considering that and assuming that the two stellar nuclei in NGC 908 contain SMBHs, we conclude that the observed circumnuclear star-forming ring in this galaxy is another evidence that this double stellar nucleus is the result of a minor merger, probably with a considerably high mass ratio. It is important to mention, however, that we do not have real evidence of the presence of SMBHs in the stellar nuclei in NGC 908 (as we did not detect AGNs in these nuclei). Even so, the observed stellar clusters may play a similar role, which, therefore, could result in the detected circumnuclear H II regions. Other examples of galaxies with circumnuclear star-forming rings for which the scenario involving a minor merger was invoked are NGC 278 \citep{kna04}, NGC 3310 \citep{smi96}, and NGC 7742 \citep{maz06}. It is worth emphasizing that the circumnuclear rings in these three galaxies have diameters of 1 - 2 kpc; therefore, the star-forming structure detected in this work (with dimensions of 200 pc $\times$ 100 pc) is significantly smaller.

An interesting characteristic in the data cube of NGC 908 is the morphological similarity between the integrated flux images of the [N \textsc{II}] $\lambda6584$, [S \textsc{II}] $\lambda6717+\lambda6731$, and H$\alpha$ emission lines (Figure~\ref{fig3}) and the $\sigma_{gas}$ and $\sigma_*$ maps, with the $\sigma_{gas}$ and $\sigma_*$ values being lower along the asymmetric star-forming ring. This indicates that the stars in this ring are forming from cold gas (with low $\sigma_{gas}$) and are keeping the low values of the velocity dispersion. Since the emission and, therefore, the velocity dispersion of young stellar populations dominate the stellar spectrum, the measured values of $\sigma_*$ along the ring are also low \citep{all05,all06}.

\subsection{NGC 1187}

The fact that the values of $T_{med}$ and $Z_{med}$ obtained for P1 and P2 in the data cube of NGC 1187 are compatible, at the 1$\sigma$ level, together with the visible similarities between the detected stellar populations (Figure~\ref{fig8}) in these regions, does not suggest that such stellar concentrations are the noncoalesced stellar nuclei of two galaxies involved in a merger. We applied the same procedure used for the data cube of NGC 908 and, using the Starlight software, fitted the spectrum of NGC 1187 - P2 taking into account only the stellar populations detected in the spectrum of NGC 1187 - P1. The $\chi^2$ of the new fit ($\chi^2_{P2}(base~P1) = 2429$) was very similar to the $\chi^2$ of the original fit ($\chi^2_{P2}=2423$). Similar results were obtained by fitting the spectrum of NGC 1187 - P1 considering only the stellar populations detected in NGC 1187 - P2. The obtained $\chi^2$ ($\chi^2_{P1}(base~P2) = 2259$) was similar to the original value ($\chi^2_{P1} = 2251$). Such results confirm the similarities between the stellar populations in NGC 1187 - P1 and in NGC 1187 - P2. However, since the double stellar nuclear structure in this galaxy is more compact than that in NGC 908, one possibility is that P1 and P2 are indeed the stellar nuclei of two merging galaxies but much closer to the final coalescence than in the case of NGC 908. Hence, due to the proximity of P1 and P2, their stellar populations are already considerably mixed, which results in the observed similarities and in compatible values for $T_{med}$ and $Z_{med}$. Although improbable, the hypothesis that the stellar nuclei of the merging galaxies have stellar populations with similar ages and metallicities must also be considered. The integrated flux images of the [N \textsc{II}] $\lambda6584$, [S \textsc{II}] $\lambda6717+\lambda6731$, and H$\alpha$ emission lines in the data cube of NGC 1187 (Figure~\ref{fig3}) have morphologies similar to that of the stellar data cube of this galaxy (Figure~\ref{fig2}), with a double nucleus, and do not show circumnuclear structures characteristic of star-forming regions (like the circumnuclear asymmetric emitting ring observed in the data cube of NGC 908). On the other hand, as mentioned before, depending on the mass ratio, a minor merger between two galaxies may not result in a significant nuclear starburst \citep{tan96,cox08}. Considering all of this, we conclude that the occurrence of a merger is not the most likely scenario to explain the double stellar and gas nucleus in the data cube of NGC 1187, but it cannot be discarded.

The position and morphology of the central disturbance (within a radius of $\sim 1\arcsec\!\!.5$ from the kinematic center) in the $V_{gas}$ map of NGC 1187 (Figure~\ref{fig10}) suggest that such structure may be related to the double nucleus in this galaxy. The behavior of the $V_{gas}$ curve in the central region corresponding to the disturbance is consistent with a rotating nuclear gas disk. In addition, the $V_*$ map and curve (Figure~\ref{fig10}) indicate the presence of a rotating nuclear stellar disk in the same area of the nuclear gas disk. As mentioned before, the presence of a nuclear eccentric stellar disk can result in a double galactic nucleus (the most famous example being M31). Therefore, we conclude that a plausible scenario to explain the double stellar and gas nucleus in NGC 1187 involves the existence of an eccentric stellar and gas rotating nuclear disk. The kinematic maps and curves in Figure~\ref{fig10} do not show features typically seen in rotating eccentric disks, like asymmetries in the radial velocity maps or an off-centered peak in the velocity dispersion maps (e.g. Tremaine 1995; Peiris \& Tremaine 2003; Salow \& Statler 2004). However, the absence of such features does not invalidate the scenario with an eccentric disk, which can also result in symmetric velocity maps or different positions for the peak of the velocity dispersion map, for example, depending on the orientation of the disk. 

The $V_{gas}$ curve of NGC 1187 shows a change in the kinematic behavior in areas located farther than $\sim 1\arcsec\!\!.5$ from the kinematic center. One possible interpretation for this is that the nuclear rotating gas disk is superimposed on a larger rotating gas disk. The same superposition may also occur for the stellar kinematics; however, we cannot evaluate that, as we were not able to measure the stellar kinematics at farther distances from the kinematic center (due to the low values of the $S/N$ ratio). In order to analyze in more detail the apparent superposition observed in the gas kinematics, we tried to decompose the rotating structures into a nuclear (possibly a rotating eccentric disk) and a circumnuclear component. This procedure was performed by fitting the H$\alpha$ emission line in each spectrum of the data cube, after the starlight subtraction, with a sum of two Gaussian functions, representing the emissions from the nuclear and circumnuclear disks. However, we observed a lot of degeneracies and also instabilities in this fitting process, possibly due to the very small widths of the emission lines in this galaxy. As a consequence, the obtained results were not sufficiently reliable to be included in this work.

In order to evaluate the consistency of the eccentric nuclear disk model with the observed spatial morphology in the central region of NGC 1187, we compared the integrated flux image of the stellar data cube of this galaxy (Figure~\ref{fig2}) with an analogous image from a data cube of the central region of M31 (obtained with the same instrumental configuration used for the observations of NGC 908 and NGC 1187). The comparison of these images is shown in Figure~\ref{fig11}. We can see that, despite the different spatial dimensions, the morphologies are very similar, which supports the hypothesis of the eccentric nuclear disk model. One point against such a model is that stable long-lived eccentric disks only exist within the radius of influence of the SMBH. Eccentric disks at larger scales naturally evolve into circular disks. It seems unlikely that an SB(r)c galaxy like NGC 1187 has an SMBH large enough so that its radius of influence extends to a scale of $\sim 100$ pc, which is the approximate size of the possible nuclear stellar disk in this galaxy. Although this fact is not sufficient to rule out the eccentric disk hypothesis, alternative scenarios must be considered. One possibility is that the double nucleus stellar and gas nucleus in NGC 1187 is the result of a more common circular stellar and gas disk, subject to perturbations that introduced distortions in the structure, originating the apparent double nucleus. 

Therefore, based on the observed stellar and gas kinematics (although a reliable decomposition of the nuclear and circumnuclear gas kinematic components was not possible) and also on the observed spatial morphology, we conclude that the most likely scenario to explain the double stellar and gas nucleus in NGC 1187 involves the stellar and gas kinematics, possibly in the form of a circular disk subject to perturbations.

\section{Conclusions}

We analyzed the stellar and gas content and the stellar and gas kinematics in the data cubes of the central regions of NGC 908 and NGC 1187, obtained with GMOS/IFU. The main conclusions of our work are as follows: 

\begin{itemize}

\item The spatial morphologies of the stellar emission in both data cubes are consistent with double stellar nuclei. In the case of the data cube of NGC 1187, the spatial morphology of the line-emitting regions is also consistent with a double nucleus. On the other hand, the main line-emitting regions in the central region of NGC 908 form a circumnuclear asymmetric ring.

\item The emission-line ratios of the nuclear and circumnuclear spectra in both data cubes are consistent with those of H II regions. In particular, similar line ratios can be observed along the circumnuclear asymmetric ring in the data cube of NGC 908, which can be identified as a star-forming ring.

\item The spectral synthesis revealed very similar stellar populations in the two stellar nuclei of NGC 1187, as the values of $T_{med}$ and $Z_{med}$ obtained for them are compatible, at the 1$\sigma$ level. On the other hand, the spectral synthesis detected differences in the stellar populations in the two stellar nuclei of NGC 908, as the values of $T_{med}$ determined for these nuclei are not compatible, even at the 3$\sigma$ level.

\item The kinematic maps of NGC 908 suggest the presence of a gas and stellar rotating disk in the central region of this galaxy. The PAs of the lines of nodes of the $V_{gas}$ and $V_*$ maps are compatible, at the 1$\sigma$ level, but are not compatible with $PA_{P1-P2}$, even at the 3$\sigma$ level, which suggests that the origin of the double stellar nucleus in this object is not related to the stellar kinematics. 

\item The $\sigma_{gas}$ and $\sigma_*$ maps of NGC 908 reveal that the velocity dispersion values are lower along the area corresponding to the circumnuclear asymmetric star-forming ring, indicating that the stars in this region are forming from cold gas and are keeping the low values of the velocity dispersion.

\item The most likely scenario to explain the double stellar nucleus in NGC 908 involves the occurrence of a minor merger, probably with a high mass ratio. A smaller (possibly satellite) galaxy being cannibalized is a possibility. Such an event could result in circumnuclear star-forming regions, like the asymmetric ring observed in this galaxy.

\item The kinematic maps of NGC 1187 reveal a rotation pattern, but with some peculiar features. The $V_{gas}$ map and curve suggest the superposition of a nuclear (within a radius of $1\arcsec\!\!.5$) and a circumnuclear gas disk. An apparent nuclear stellar disk, nearly cospatial with the nuclear gas disk, can also be seen in the $V_*$ map. The PAs of the lines of nodes of the $V_{gas}$ and $V_*$ maps are compatible, at the 1$\sigma$ level, and both are compatible with $PA_{P1-P2}$, at the 1$\sigma$ level. 

\item The most likely scenario to explain the double stellar and gas nucleus in NGC 1187 involves the stellar and gas kinematics, in the form of a circular disk subject to perturbations.

\end{itemize}

\acknowledgments

Based on observations obtained at the Gemini Observatory (processed using the Gemini IRAF package), which is operated by the Association of Universities for Research in Astronomy, Inc., under a cooperative agreement with the NSF on behalf of the Gemini partnership: the National Science Foundation (United States), the National Research Council (Canada), CONICYT (Chile), the Australian Research Council (Australia), Minist\'{e}rio da Ci\^{e}ncia, Tecnologia e Inova\c{c}\~{a}o (Brazil) and Ministerio de Ciencia, Tecnolog\'{i}a e Innovaci\'{o}n Productiva (Argentina). This work was supported by CAPES (PNPD - PPG-F\'isica - UFABC) and FAPESP (under grant 2011/51680-6). We thank the anonymous referee for valuable comments on the paper.

{\it Facilities:} \facility{Gemini:South(GMOS)}.

\end{document}